\documentclass{iaus}
\usepackage{graphicx}

\title[Irregular Satellites] 
{Outer Irregular Satellites of the Planets and Their Relationship with Asteroids, Comets and Kuiper Belt Objects}

\author[Scott Sheppard]   
{Scott S. Sheppard$^1$}

\affiliation{$^1$Department of Terrestrial Magnetism, Carnegie Institution of Washington,
Washington, DC 20015, USA \break email: sheppard@dtm.ciw.edu \\}

\pubyear{2006}
\volume{229}  
\pagerange{100--120}
\date{?? and in revised form ??}
\jname{Asteroids, Comets, Meteors}
\editors{D. Lazzaro, S. Ferraz-Mello \& J. Fernandez, eds.}
\begin{document}

\maketitle

\begin{abstract}

Outer satellites of the planets have distant, eccentric orbits that
can be highly inclined or even retrograde relative to the equatorial
planes of their planets. These irregular orbits cannot have formed by
circumplanetary accretion and are likely products of early capture
from heliocentric orbit.  The irregular satellites may be the only
small bodies remaining which are still relatively near their formation
locations within the giant planet region.  The study of the irregular
satellites provides a unique window on processes operating in the
young solar system and allows us to probe possible planet formation
mechanisms and the composition of the solar nebula between the rocky
objects in the main asteroid belt and the very volatile rich objects
in the Kuiper Belt.  The gas and ice giant planets all appear to have
very similar irregular satellite systems irrespective of their mass or
formation timescales and mechanisms.  Water ice has been detected on
some of the outer satellites of Saturn and Neptune whereas none has
been observed on Jupiter's outer satellites.

\end{abstract}

\firstsection 
\section{Introduction}

Satellites are stable in the region called the Hill sphere in which
the planet, rather than the sun, dominates the motion of the object
(Henon 1970).  The Hill sphere radius of a planet is defined as
\begin{equation}
r_H = a_p \left[\frac{m_p}{3M_{\odot}}\right]^{1/3}
\label{planethill}
\end{equation}
where $a_p$ and $m_p$ are the semi-major axis and mass of the planet
and $M_{\odot}$ is the mass of the sun.  Table \ref{tab:planetinfo} shows
the sizes of each giant planet's Hill sphere as seen from the Earth at
opposition.

Most planetary satellites can be classified into one of two
categories: regular or irregular (Kuiper (1956); \cite{Peale99}).

The regular satellites are within about $0.05 r_{H}$ and have nearly
circular, prograde orbits with low inclinations near the equator of
the planet.  These satellites are thought to have formed around their
respective planets through circumplanetary accretion, similar to how
the planets formed in the circumstellar disk around the sun.  The
regular satellites can be subdivided into two types: classical
regulars and collisional shards (Burns 1986).

The classical regular satellites are large (several hundred to
thousands of kilometers in size) and have evenly spaced orbits.  The
regular collisional shards are small (less than a few hundred
kilometers) and are believed to have once been larger satellites but
have been shattered or tidally disrupted over their lifetimes.  These
shards are usually very near the planet where tidal forces and meteor
fluxes are very high.  Many collisional shards are associated with
known planetary rings.

\begin{figure}
\centerline{\includegraphics[height=5in,angle=0]{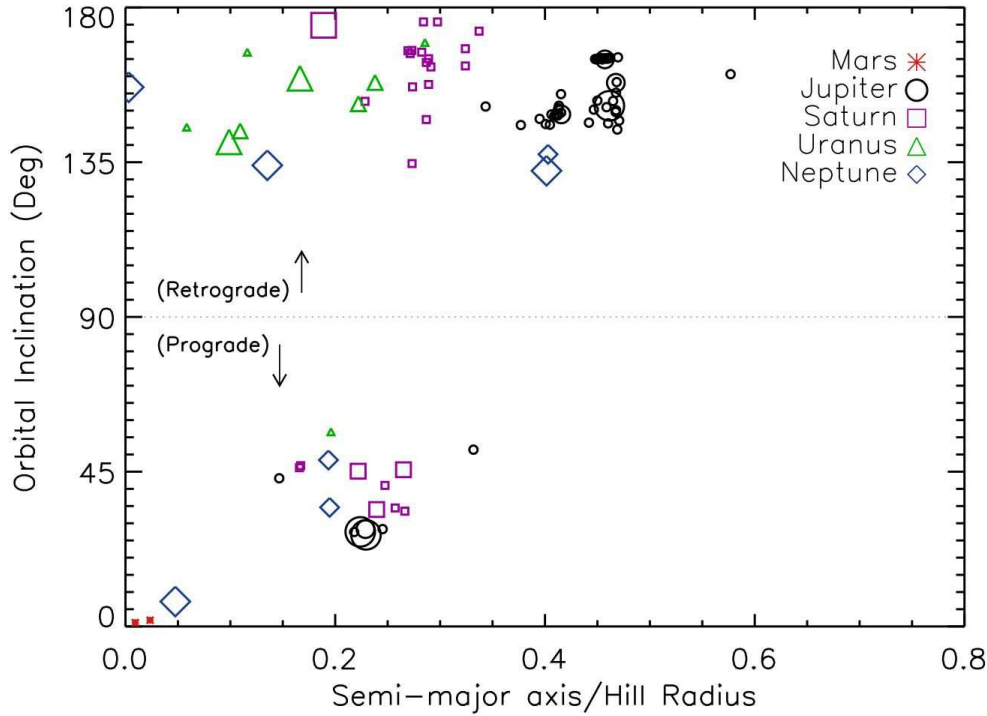}}
\caption{All 96 Known irregular satellites of the giant planets.  The
horizontal axis is the ratio of the satellites semi-major axis to the
respective planet's Hill radius.  The vertical axis is the inclination
of the satellite to the orbital plane of the planet.  The size of the
symbol represents the radius of the object: Large symbol $r > 25$ km,
medium symbol $25 > r > 10$ km, and small symbol $r<10$ km.  Neptune's
Triton can be seen in the upper left of the figure while Nereid is
near the lower left.  Mars' two satellites are plotted for comparison.
All 53 known regular satellites would fall near the origin of this
plot. [Modified from Sheppard et al. 2005]}
\label{fig:allirr}
\end{figure}

In contrast, the irregular satellites have semi-major axes $> 0.05
r_{H}$ with apocenters up to $0.65 r_{H}$ (Figure \ref{fig:allirr}).
Irregular satellites have eccentric orbits that are usually highly
inclined and distant from the planet.  They can have both prograde and
retrograde orbits.  The irregular satellites can not have formed
around their respective planet with their current orbits and are
likely the product of early capture from heliocentric orbits
(Kuiper 1956).

\begin{table}
  \begin{center}
  \caption{Irregular Satellites of the Planets}
  \label{tab:planetinfo}
  \begin{tabular}{lcccccc}\hline
     Planet  & Irr$^{a}$  & $m_{p}$        & $r_{min}$$^{b}$ & $a_{\mbox{crit}}$ & $r_{H}$$^{c}$ & $r_{H}$ \\
             & (\#) & ($10^{25}$kg)  & (km)      & ($10^{6}$km)            & (deg)   & ($10^{7}$km) \\\hline
     Mars    & 0    &  0.06          &  0.1      &                   & 0.8     &  0.1  \\
     Jupiter & 55   &  190           &  1.5      & 6.6               & 4.7     &  5.1  \\
     Saturn  & 26   &  57            &  3        & 5.7               & 3.0     &  6.9  \\
     Uranus  & 9    &  9             &  7        & 2.9               & 1.5     &  7.3  \\
     Neptune & 6(7) &  10            &  16       & 3.8               & 1.5     &  11.6  \\\hline
  \end{tabular}
 \end{center}
a) The number of known irregular satellites. \\
b) Minimum radius that current outer satellite searches would have detected to date. \\
c) Size of the Hill sphere as seen from Earth at opposition.
\end{table}

\begin{figure}
\centerline{\includegraphics[height=5in,angle=0]{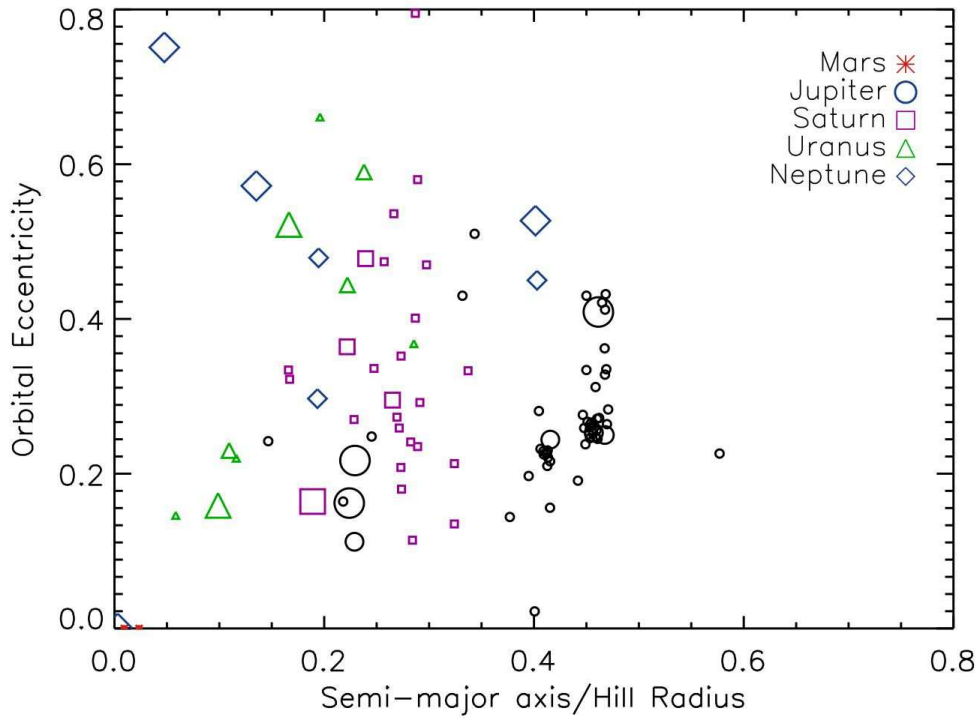}}
\caption{All 96 Known irregular satellites of the giant planets.  The
horizontal axis is the ratio of the satellites semi-major axis to the
respective planet's Hill radius.  The vertical axis is the orbital
eccentricity.  The size of the symbol represents the radius of the
object: Large symbol $r > 25$ km, medium symbol $25 > r > 10$ km, and small
symbol $r < 10$ km.  Again, all 53 known regular satellites would fall
near the origin of this plot, where Triton and Mars' satellites are
located. [Modified from Sheppard et al. 2005]}
\label{fig:allirrae}
\end{figure}

Orbital characteristics displaying strong capture signatures instead
of in situ formation around the planet is often used to define an
irregular satellite.  Throughout this work we will use a more strict
definition.  We follow others and define irregular satellites as those
satellites which are far enough from their parent planet that the
precession of their orbital plane is primarily controlled by the sun
instead of the planet's oblateness.  In other words, the satellite's
inclination is fixed relative to the planet's orbit plane instead of
the planet's equator.  In practice this means any satellite with a
semi-major axis more distant than the critical semi-major axis
(Burns 1986), $a_{\mbox{crit}} \sim (2 J_{2} r_{p}^{2}
a_{p}^{3} m_{p} / M_{\odot} )^{1/5}$, is an irregular satellite (Table
\ref{tab:planetinfo}).  Here $J_{2}$ is the planet's second gravitational
harmonic coefficient and $r_{p}$ is the planet's equatorial radius.
Figures \ref{fig:allirr} and \ref{fig:allirrae} show the orbital
characteristics of the known irregular satellites.  In these figures
all known regular satellites would fall very near the origin.

Almost all known planetary satellites fall into one of the three types
mentioned above.  A few exceptions do exist.  The formation of the
Earth's Moon is best explained through a collision between a Mars
sized body and the young Earth (see Canup \& Asphaug 2001 and
references therein).  Mars' two small satellites Phobos and Deimos
resemble regular collisional shards, but some have suggested that they
may be captured bodies similar to the irregular satellites of the
giant planets (Burns 1992).  No outer irregular satellites of Mars are
known to exist (Sheppard \& Jewitt 2004).  Both Neptune's Triton and
Saturn's Iapetus have at times been considered irregular satellites.
These two objects stand out because both Triton and Iapetus are about
ten times larger than any other known irregular satellite.  Triton has
all the characteristics of a regular satellite except that its orbit
is retrograde.  The best explanation for a retrograde orbit is through
capture.  Iapetus also has all the characteristics of a regular
satellite but its inclination of 7 degrees is significantly larger
than any other known regular satellite.  Even so, this inclination is
not as high as the vast majority of irregular satellites.  Iapetus'
relatively large inclination is probably because it is the most
distant regular satellite of Saturn.  At this distance the
circumSaturnian nebula was probably of low density which dissipated
quickly.  These factors would have significantly slowed or stopped the
process of orbital evolution.  Finally, Neptune's outer satellite
Nereid is usually considered an irregular satellite but its relatively
small semi-major axis and low inclination yet exceptionally large
eccentricity suggest it may be a perturbed regular satellite, perhaps
from Triton's capture (Goldreich et al. 1989; Cuk \& Gladman 2005;
Sheppard et al. 2006).

\section{Irregular Satellite Discovery}

\begin{figure}
\centerline{\includegraphics[height=4in,angle=90]{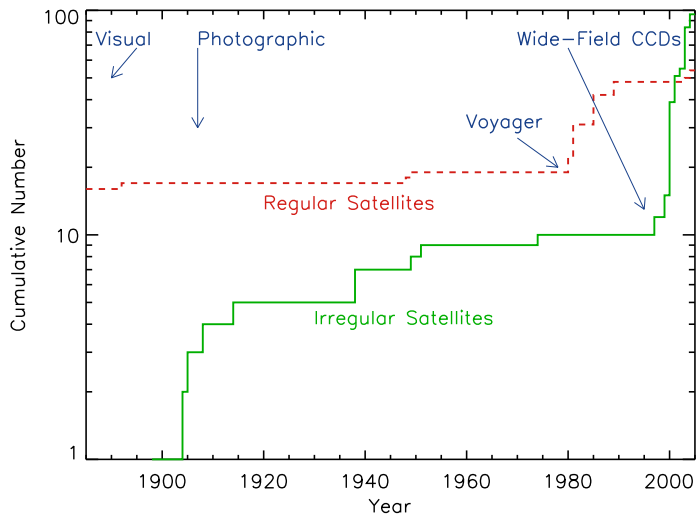}}
\caption{The number of irregular and regular satellites discovered
since the late 1800's.  Key technological advances which resulted in a
jump in discoveries are listed.}
\label{fig:discsats}
\end{figure}

Irregular satellite discovery requires large fields of view because of
the large planetary Hill spheres.  Sensitivity is needed because the
majority of irregular satellites are small (radii $< 50$ km) and
therefore faint.  With the use of large field-of-view photographic
plates around the end of the 1800's the first distinctive irregular
satellites were discovered (Figure \ref{fig:discsats}).  In 1898 the
largest irregular satellite of Saturn, Phoebe (radius $\sim 60$ km),
was discovered and in 1904 the largest irregular of Jupiter, Himalia
(radius $\sim 92$ km), was discovered (see Kuiper 1961 for a review of
early photographic surveys).

Until 1997 only ten or eleven irregular satellites were known and the
last discovered irregular satellite was in 1975 on photographic plates
(Kowal et al. 1975).  Since 1997 eighty-six irregular satellites have
been discovered around the giant planets (Gladman et
al. 1998;2000;2001; Sheppard \& Jewitt 2003; Holman et al. 2004;
Kavelaars et al. 2004; Sheppard et al. 2005;2006).  Jupiter's retinue
of irregular satellites has increased from 8 to 55, Saturn's from 1 to
26, Uranus' from 0 to 9 and Neptune's from 1 to 6 (or seven if
including Triton).  Table \ref{tab:planetinfo} shows information about
the current irregular satellite systems around the giant planets.

\begin{figure}
\centerline{\includegraphics[height=4in,angle=0]{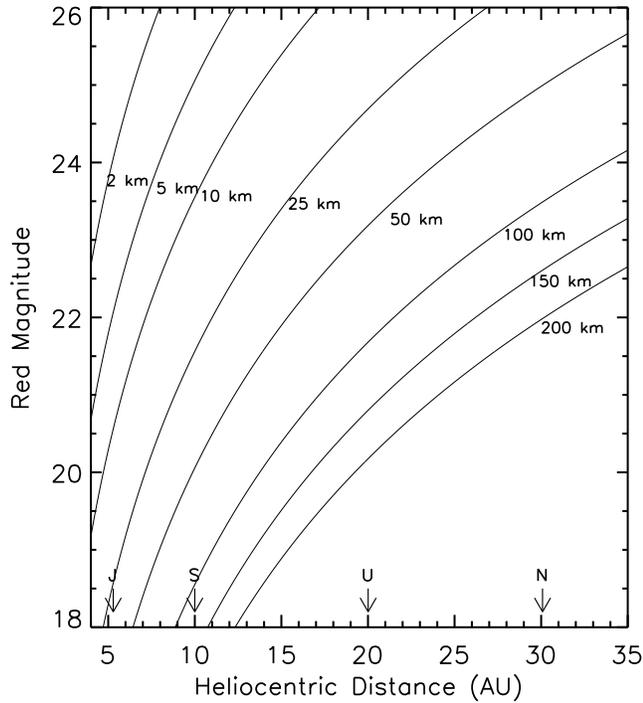}}
\caption{The distances of the planets versus the observable small body
population diameter for a given red magnitude assuming an albedo of
0.04.  Jupiter's closer proximity allows us to probe the smallest
satellites.}
\label{fig:distance26}
\end{figure}

The number of known irregular satellites (96 as of November 2005) have
recently surpassed the number of known regular satellites (Figure
\ref{fig:discsats}).  The main reason is new technology.  The recent
development of sensitive, large scale CCD detectors has allowed these
faint outer planetary satellites to be discovered.  Because of the
proximity of Jupiter (Figure \ref{fig:distance26}), it currently has
the largest irregular satellite population (Sheppard and Jewitt 2003).

\section{Capture of Irregular Satellites}

\begin{figure}
\centerline{\includegraphics[height=4in,angle=0]{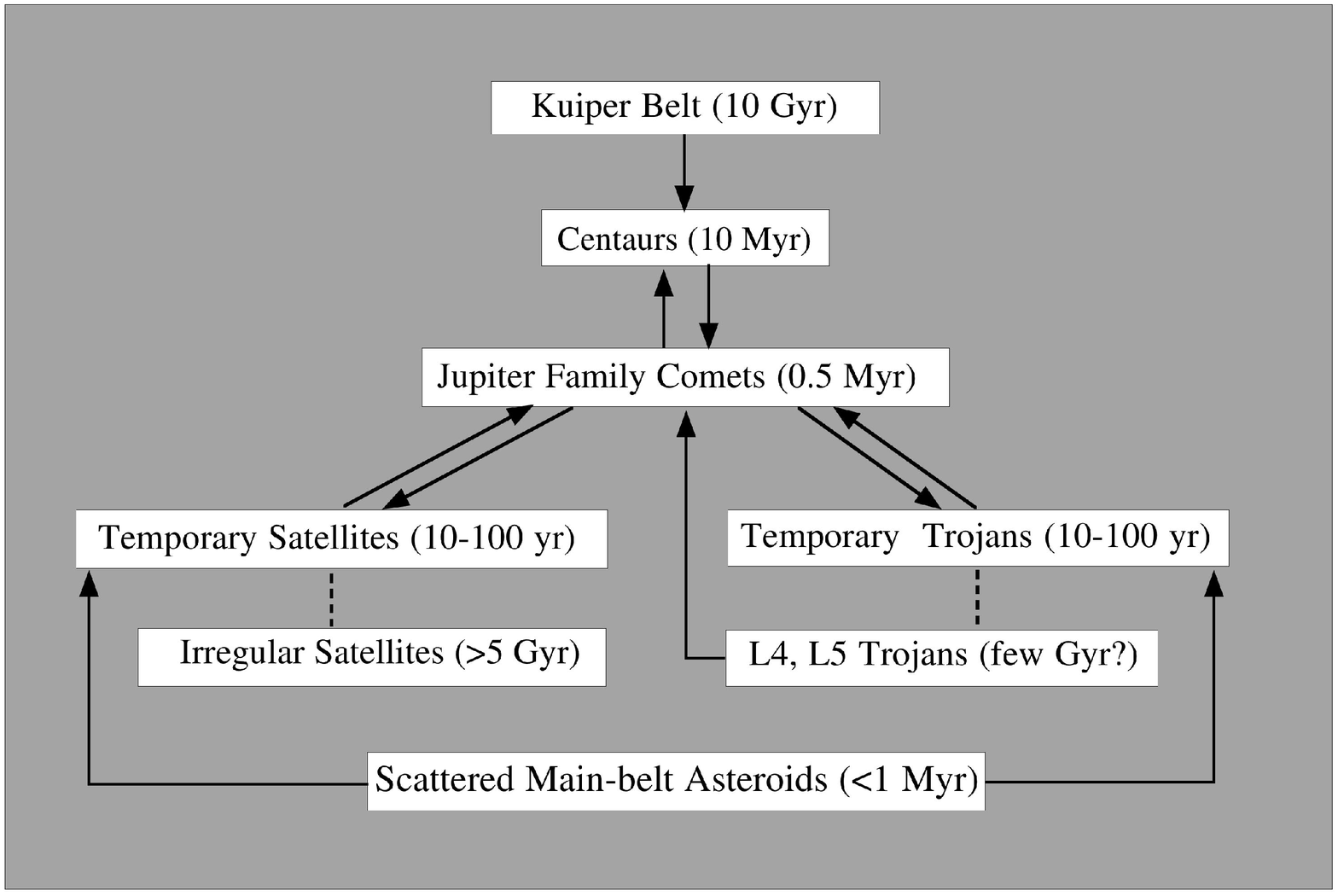}}
\caption{Interrelations among the small body populations in the solar
system.  Solid arrows denote established dynamical pathways.  Dashed
lines show pathways which currently have no known energy dissipation
source and thus can not lead to permanent capture but only temporary
capture.  During the planet formation epoch such pathways may have
existed.  Numbers in parentheses indicate the approximate dynamical
lifetimes of the different populations. (Figure from Jewitt et
al. (2004))}
\label{fig:capturepaths}
\end{figure}

Only the four giant planets have known irregular satellite populations
(Figure \ref{fig:allirr}).  The likely reason is that the capture
process requires something that the terrestrial planets did not have.
Capture of a heliocentric orbiting object is likely only if the object
approaches the planet near its Lagrangian points and has an orbital
velocity within about $1 \%$ that of the planet.  Objects may
temporarily orbit a planet (i.e. Shoemaker-Levy 9) but because of the
reversibility of Newton's equations of motion some form of energy
dissipation is required to permanently capture a body.  Without
dissipation the object will be lost in less than a few hundred years
(Everhart 1973; Heppenheimer \& Porco 1977).  In the present epoch a planet has
no known efficient mechanism to permanently capture satellites (Figure
\ref{fig:capturepaths}).

Kuiper (1956) first suggested that the irregular satellites were
originally regular satellites which escaped from the planet's Hill
sphere to heliocentric orbit because of the decreasing mass of the
planet.  These ``lost'' satellites would have similar orbits as the
parent planet.  Eventually the satellite would pass near the planet
and be slowed down from the mass escaping from the planet.  The
satellite would thus be captured in an ``irregular'' type orbit.  It
is now believed that the giant planets never lost significant amounts
of mass and thus irregular satellites are unlikely to be escaped
regular satellites.

The dissipation of energy through tidal interactions between the
planet and irregular satellites is not significant for such small
objects at such large distances (Pollack et al. 1979).  The creation
of irregular satellites from explosions of the outer portions of the
massive ice envelopes of the large regular satellites from saturation
by electrolysis seems unlikely and no observational evidence supports
such explosions on the regular satellites (Agafonova \& Drobyshevski
1984).

Three viable mechanisms have been proposed for irregular satellite
capture.  Satellite capture could have occurred efficiently towards
the end of the planet formation epoch due to gas drag from an extended
planetary atmosphere (Kuiper 1956; Pollack, Burns \& Tauber 1979), the
enlargement of the Hill sphere caused by the planet's mass growth
(Heppenheimer \& Porco 1977) and/or higher collisional or
collisionless interaction probabilities with nearby small bodies
(Colombo \& Franklin 1971; Tsui 2000).  Below we discuss each of these
in more detail.

\subsection{Capture by Gas Drag}

During early planet formation the giant planets likely had primordial
circumplanetary nebulae (Pollack et al. 1979; Cuk \& Burns 2004).  An
object passing through this gas and dust near a planet would have
experienced gas drag.  In order to significantly slow an object for
capture it would need to encounter about its own mass within the
nebula.  Conditions at the distances of the irregular satellites are
unknown, but rough estimates suggest that if the object was larger
than a few hundred kilometers it would not have been significantly
affected.  If the object was very small it would have been highly
slowed and would have spiraled into the planet.  If the object was
just the right size (a few km to a few hundred kilometers) is would
have experienced just enough gas drag to be captured (Pollack et
al. 1979).  Hydrodynamical collapse of the primordial planetary nebula
would have to occur within a few thousand years of capture in order
for the satellites to not experience significant orbital evolution and
eventually spiral into the planet from gas drag.  In this scenario the
current irregular satellites are only the last few captured bodies
which did not have time to spiral into the planet.  Retrograde objects
would have experienced larger gas drag during their time within the
nebula and thus their orbits should be more modified toward smaller
eccentricities, inclinations and semi-major axes.  Observations
currently show that both the progrades and retrogrades have similar
modification.  Gas drag would also allow for larger objects to be
captured closer to the planet since the nebula would be more dense
there.  In the action of gas drag smaller irregular satellites should
have their orbits evolve faster and should have been preferentially
removed.  No size versus orbital characteristics are observed for any
of the irregular satellites of the planets.

\subsection{Pull-down Capture}

Another way an object can become permanently captured is if the
planet's mass increased or the Sun's mass decreased while the object
was temporarily captured, called pull-down capture (Heppenheimer \&
Porco 1977).  Either of these scenarios would cause the Hill sphere of
the planet to increase making it impossible for the object to escape
with its current energy.  Again, the enlargement of the Hill sphere
would have to happen over a short timescale.  Likely mass changes of
the Sun or the planet would need to be greater than about $40 \%$ over
a few thousand years (Pollack et al. 1976).  The Hill sphere of the
planet would also increase if the planet migrated significantly away
from the sun (Brunini 1995). This mechanism is not a likely cause of
permanent capture because the large migrations required to make
temporary capture permanent within a few thousand years would severely
disrupt any satellite systems (Beauge et al. 2002).

\subsection{Capture Through Collisional or Collisionless Interactions}

Finally, a third well identified mechanism of capture could be from
the collision or collisionless interaction of two small bodies within
the Hill sphere of the planet (Colombo \& Franklin; Tsui 2000;
Astakhov et al. 2003; Funato et al. 2004; Agnor \& Hamilton 2004).
This could occur as asteroid-asteroid or asteroid-satellite
encounters.  These encounters could dissipate the required amount of
energy from one or both of the objects for permanent capture.  This
mechanism for capture would operate much more efficiently during the
early solar system when many more small bodies where passing near the
planets.  An interesting point of this capture mechanism is that it
would be fairly independent of the mass or formation scenario of the
planet and mostly depend on the size of the Hill sphere and number of
passing bodies.

\section{Dynamics of Irregular Satellites}

The known irregular satellites are stable over the age of the Solar
System though strongly influenced by solar and planetary perturbations
(Henon 1970; Carruba et al. 2002; Nesvorny et al. 2003).  The
perturbations are most intense when the satellite is near apoapsis.
High inclination orbits have been found through numerical simulations
to be unstable due to solar perturbations (Carruba et al. 2002;
Nesvorny et al. 2003).  Satellites with inclinations between $50 < i <
130$ degrees slowly have their orbits stretched making them obtain
very high eccentricities.  The high eccentricities are obtained in
$10^{7} - 10^{9}$ years and cause the satellite to eventually be lost
from the system either through exiting the Hill sphere or colliding
with a regular satellite or the planet.

A number of irregular satellites have been found to be in orbital
resonances with their planet.  These resonances protect the satellites
from strong solar perturbations.  The two main types of resonance
found to date are Kozai resonances and secular resonances (Kozai 1962;
Carruba et al. 2002; Nesvorny et al. 2003).  The irregular satellites
known or suspected of being in resonances are Jupiter's irregular
satellites Sinope, Pasiphae, Euporie (S/2001 J10), S/2003 J18 and
Carpo (S/2003 J20) and Saturn's irregular satellites Siarnaq (S/2000
S3), Kiviuq (S/2000 S5) and Ijiraq (S/2000 S6) and Uranus' Stephano
(Saha \& Tremaine 1993; Whipple \& Shelus 1993; Nesvorny et al. 2003;
Cuk \& Burns 2004; R. Jacobson person communication).  These
resonances occupy a very small amount of orbital parameter space.  The
evolution of satellites into these resonances implies some sort of
slow dissipation mechanism which allowed the satellites to acquire the
resonances and not jump over them.  This could be obtained from weak
gas drag, a small increase in the planet's mass or a slow migration of
the planet.

From numerical and analytical work it has been found that retrograde
orbits are more stable than prograde orbits over large time-scales
(Moulton 1914; Hunter 1967; Henon 1970; Hamilton \& Krivov 1997).
Analytically the retrogrades may be stable up to distances of $\sim
0.7 r_{H}$ while progrades are only stable up to $0.5 r_{H}$ (Hamilton
\& Krivov 1997).  This is consistent with known orbits of retrogrades
and progrades to date.  Known retrograde (prograde) irregular
satellites have semi-major axes out to $\sim 0.47 r_{H}$ ($\sim 0.33
r_{H}$) and have apocenters up to $\sim 0.65 r_{H}$ ($ \sim 0.47
r_{H}$).

Many of the irregular satellites have been found to show dynamical
groupings (Gladman et al. 2001; Sheppard \& Jewitt 2003; Nesvorny et
al. 2003).  At Jupiter the dynamical groupings are well observed in
semi-major axis and inclination phase space (Figure \ref{fig:allirr})
and are probably similar to families found in the main belt asteroids
which are created when a larger parent body is disrupted into several
smaller daughter fragments.  The irregular satellites at the other
giant planets are mostly grouped in inclination phase space and not in
semi-major axis phase space.  It would be unlikely that a fragmented
body would create daughter bodies with such significantly different
semi-major axes.  This inclination clustering may just be because of
resonance effects or that these particular inclinations are more
stable.  Still, there do appear to be some irregular satellites at
Saturn, Uranus and Neptune that do cluster in semi-major axis and
inclination phase space like those seen at Jupiter but these groups
are not well populated.  Further satellites in these putative
dynamical families may be observed when smaller satellites are able to
be discovered in the future.

Fragmentation of the parent satellites could be caused by impact with
interplanetary projectiles (principally comets) or by collision with
other satellites.  Collisions with comets are improbable in the
current solar system but during the heavy bombardment era nearly 4.5
billion years ago they would have been highly probable (Sheppard \&
Jewitt 2003).  Large populations of now defunct satellites could also
have been a collisional source in creating the observed satellite
groupings (Nesvorny et al. 2004).  No size versus orbital property
correlations are seen in the groupings which suggest breakup occurred
after any significant amounts of gas were left.

The detection of dust in bound orbits about Jupiter in the outer
Jupiter system from the Galileo spacecraft is attributed to high
velocity impacts of interplanetary micrometeoroids into the
atmosphereless outer satellites (Krivov et al. 2002).  The micron
sized dust is in prograde and retrograde orbits with a number density
($10$ km$^{-3}$) about ten times larger than in the local
interplanetary medium.

\section{Physical Properties of Irregular Satellites}

Most of the space in the giant planet region of the solar system is
devoid of objects which make irregular satellites one of the only
dynamical clues as to what affected most of the mass in the solar
system 4.5 billion years ago.  Irregular satellites were likely
asteroids or comets in heliocentric orbit which did not get ejected
into the Oort cloud or incorporated in the planets.  They may be some
of the only small bodies remaining which are still relatively near
their formation locations within the giant planet region.  The
irregular satellite reservoirs lie between the main belt of asteroids
and Kuiper Belt which makes them a key to showing us the complex
transition between rocky objects in the main asteroid belt and the
expected very volatile rich objects in the Kuiper Belt.

\subsection{Visible and Infrared Colors}

\begin{figure}
\centerline{\includegraphics[height=4in,angle=90]{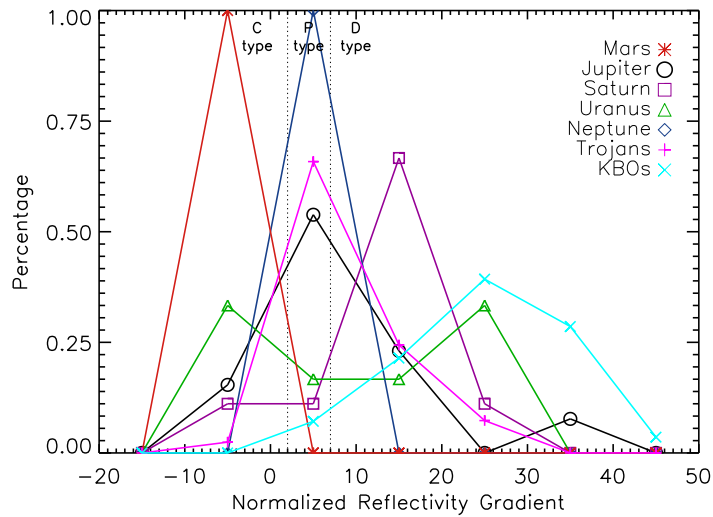}}
\caption{The colors of the irregular satellites of Jupiter, Saturn,
Uranus and Neptune compared to the KBOs, Trojans and Martian
satellites.  The Jupiter irregular satellites are fairly neutral in
color and very similar to the nearby Jupiter Trojans.  Saturn's
irregulars are significantly redder than Jupiter's but do not reach
the extreme red colors seen in the KBOs.  Uranus' irregular satellites
are very diverse in color with some being the bluest known while
others are the reddest known irregular satellites.  Only two of
Neptune's irregulars have measured colors and not much can yet be said
except they don't show the very red colors seen in the Kuiper Belt.
The general linear colors of the C, P and D-type asteroids are shown
for reference (Dahlgren \& Lagerkvist 1995). Irregular satellite
colors are from Grav et al. 2003; 2004a.}
\label{fig:colorhist}
\end{figure}

\begin{figure}
\centerline{\includegraphics[height=4in,angle=90]{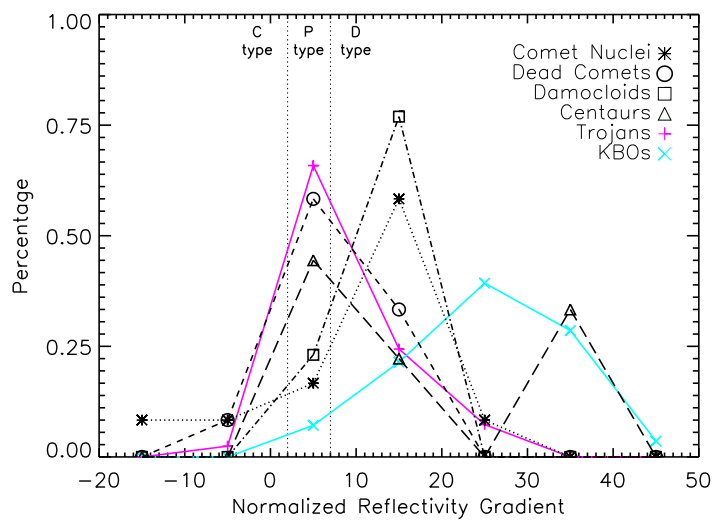}}
\caption{Same as Figure \ref{fig:colorhist} except this plot shows the
colors of the comet nuclei, dead comet candidates, Damocloids,
Centaurs, Trojans and KBOs.  It is plotted as a separate graph from
Figure \ref{fig:colorhist} to avoid confusion between the many
different types of objects.  Jupiter's irregulars are similar in color
to the dead comets and some of the Centaurs.  Saturn's irregulars are
similar in color to the Damocloids and active comet nuclei.  Comet
nuclei and dead comet colors are from Jewitt (2002) and references
therein.  Centaur and KBO colors are from Barucci et al. (2001);
Peixinho et al. (2001); Jewitt \& Luu (2001) and references therein.
Damocloid colors are from Jewitt (2005) and references therein.}
\label{fig:colorhistcom}
\end{figure}

Colors of the irregular satellites are neutral to moderately red
(Tholen \& Zellner 1984; Luu 1991; Rettig et al. 2001; Maris et
al. 2001; Grav et al. 2003; 2004a).  Most do not show the very red
material found in the distant Kuiper Belt (Figures \ref{fig:colorhist}
and \ref{fig:colorhistcom}).  The Jupiter irregular satellite colors
are very similar to the C, P and D-type carbonaceous outer main belt
asteroids (Degewij et al. 1980) as well as to the Jupiter Trojans and
dead comets.  Colors of the Jupiter irregular satellite dynamical
groupings are consistent with, but do not prove, the notion that each
group originated from a single undifferentiated parent body.  Optical
colors of the 8 brightest outer satellites of Jupiter show that the
prograde group appears redder and more tightly clustered in color
space than the retrograde irregulars (Rettig et al. 2001; Grav et
al. 2003).  Near-infrared colors recently obtained of the brighter
satellites agree with this scenario and that the Jupiter irregular's
colors are consistent with D and C-type asteroids (Sykes et al. 2000;
Grav et al. 2004b).

The Saturn irregular satellites are redder than Jupiter's but still do
not show the very red material observed in the Kuiper Belt.  The
colors are more similar to the active cometary nuclei and damocloids.
Buratti et al. (2005) show that the color of the dark side of Iapetus
is consistent with dust from the small outer satellites of Saturn.
Buratti et al. also find that none of Saturn's irregular satellites
have similar spectrophotometry as Phoebe.  The irregular satellites of
Uranus have a wide range of colors from the bluest to the reddest.
These satellites may show the extreme red colors observed in the
Kuiper Belt and have a distribution similar to the Centaurs.
Neptune's irregulars have limited observational data but to date they
don't show the extreme red colors seen in the Kuiper Belt.

\subsection{Spectra and Albedos}

Near-Infrared and optical spectra of the brightest Jupiter satellites
are mostly linear and featureless (Luu 1991; Brown 2000; Jarvis et
al. 2000; Chamberlain \& Brown 2004; Geballe et al. 2002).  Jarvis et
al. (2000) finds a possible 0.7 micron absorption feature in Jupiter's
Himalia and attributes this to oxidized iron in phyllosilicates which
is typically produced by aqueous alteration.  The spectra of Jupiter's
irregular satellites are consistent with C-type asteroids.  The
irregular satellites at Saturn and Neptune appear to be remarkably
different with rich volatile surfaces..  The largest Saturn irregular,
Phoebe, has been found to have water ice (Owen et al. 1999) as has the
large Neptune irregular satellite Nereid (Brown, Koresko \& Blake
1998).

Jupiter's irregular satellites have very low albedos of about 0.04 and
0.05 which again along with their colors are consistent with dark C, P
and D-type Carbon rich asteroids in the outer main belt (Cruikshank
1977) and very similar to the Jovian Trojans (Fernandez et al. 2003).
Saturn's Phoebe has an average albedo of about 0.07 (Simonelli et
al. 1999) while Neptune's Nereid was found to have an albedo of 0.16
from Voyager data (Thomas et al. 1991).  These albedos are more
similar to the higher albedos found in the Kuiper Belt (Grundy et
al. 2005; Cruikshank et al. 2005).  These are in comparison to the
average albedos of comet nuclei 0.03, extinct comets 0.03, and Jovian
Trojans 0.06 (Fernandez et al. 2003).

The Cassini spacecraft obtained resolved images of Himalia and showed
it to be an elongated shaped object with axes of 150 x 120 km with an
albedo of about 0.05 (Porco et al. 2003).  Cassini obtained a mostly
featureless near-infrared spectrum of Jupiter's JVI Himalia
(Chamberlain \& Brown 2004).

\begin{figure}
\centerline{\includegraphics[height=3in,angle=0]{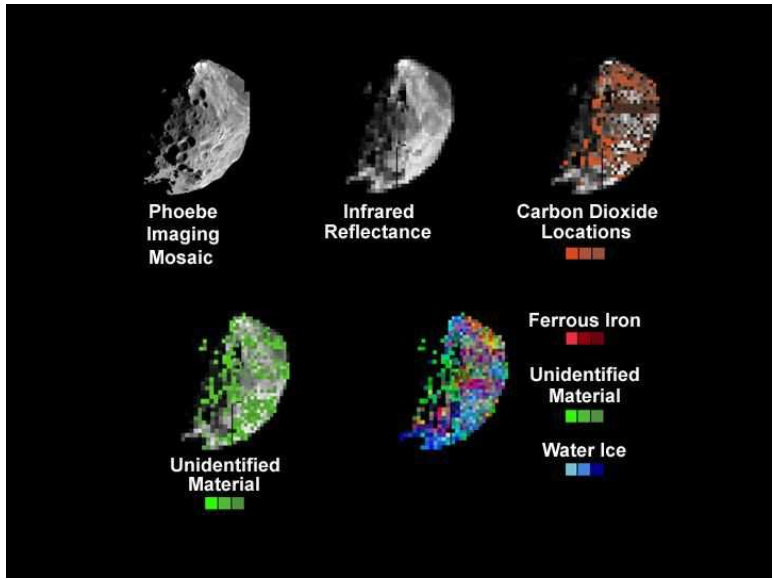}}
\caption{Phoebe's mineral distribution as seen by the Cassini
spacecraft. Phoebe appears to have a very volatile rich surface which
is unlike the irregular satellites at Jupiter.  (Produced by
NASA/JPL/University of Arizona/LPL using data from the Cassini Imager
and VIMS; see Porco et al. 2005 and Clark et al. 2005)}
\label{fig:phoebe}
\end{figure}

Cassini obtained much higher resolution images of Saturn's irregular
satellite Phoebe (Figure \ref{fig:phoebe}) with a flyby of 2071 km on
June 11, 2004.  The images showed Phoebe to be intensively cratered
with many high albedo patches near crater walls (Porco et al. 2005).
Phoebe's density was found to be $1630 \pm 33$ kg m$^{-3}$ (Porco et
al. 2005). The spectra showed lots of water ice as well as
ferrous-iron-bearing minerals, bound water, trapped CO$_{2}$,
phyllosilicates, organics, nitriles and cyanide compounds on the
surface (Clark et al. 2005).  Phoebe's volatile rich surface and many
compounds infer the object was formed beyond the rocky main belt of
asteroids and maybe very similar to the composition of comets.  Its
relatively high density compared to that observed for comets and
inferred for Kuiper Belt objects makes it a good candidate to have
formed near its current location where the highly volatile materials
are still unstable to evaporation.

\section{Comparison of the Giant Planet Irregular Satellite Systems}

\subsection{Giant Planet Formation}

Irregular satellites are believed to have been captured around the
time of the formation of the giant planets.  Thus their dynamical and
physical properties are valuable clues as to what happened during the
planet formation process.  Because of the massive hydrogen and helium
envelopes of the gas giants Jupiter and Saturn, they presumably formed
quickly in the solar nebula before the gas had time to significantly
dissipate.  The less massive and deficient in hydrogen and helium ice
giants Uranus and Neptune appear to have taken a drastically different
route of evolution.

There are two main models for giant planet formation.  The standard
model of core accretion assumes the cores of the giant planets were
formed through oligarchic growth for about $10^{6}$ to $10^{8}$ years.
Once they obtained a core of about ten Earth masses they quickly
accreted their massive gaseous envelopes (Pollack et al. 1996).  The
disadvantage of this model is that the protoplanetary disk likely
dissipated within a few million years while the core accretion model
requires long timescales to form the planets.  Because of the lower
surface density and larger collisional timescales for more distant
planets the core accretion model can not adequately form Uranus and
Neptune in the age of the solar system.

The second giant planet formation mechanism is through disk
instabilities.  This model suggests parts of the solar nebula became
unstable to gravitational collapse (Boss 2001).  In this model the
planets would form on timescales of only about $10^{4}$ years.  The
disadvantages are it doesn't allow for massive cores and does not
appear to be applicable to the small masses of Uranus and Neptune.

Both giant planet formation models have trouble forming Uranus and
Neptune (Bodenheimer \& Pollack 1986; Pollack et al. 1996).  Any
theory on the different formation scenarios of Uranus and Neptune to
that of Jupiter and Saturn should take into account the irregular
satellite systems of each.  The recent theory that Uranus and Neptune
lost their hydrogen and helium envelopes by photoevaporation from
nearby OB stars (Boss, Wetherill, \& Haghighipour 2002) would have
caused all their irregular satellites to be lost because of the
significant decrease in the planet's mass (Sheppard and Jewitt (2003);
Jewitt \& Sheppard (2005)).  Another recent theory is that Uranus and
Neptune formed in the Jupiter-Saturn region with subsequent scattering
to their current locations (Thommes, Duncan, \& Levison 2002).  Any
large migration by the planets would have disrupted any outer
satellite orbits (Beauge et al. 2002).

\subsection{Population and Size Distributions of the Irregular Satellites}

\begin{figure}
\centerline{\includegraphics[height=2.5in,angle=0]{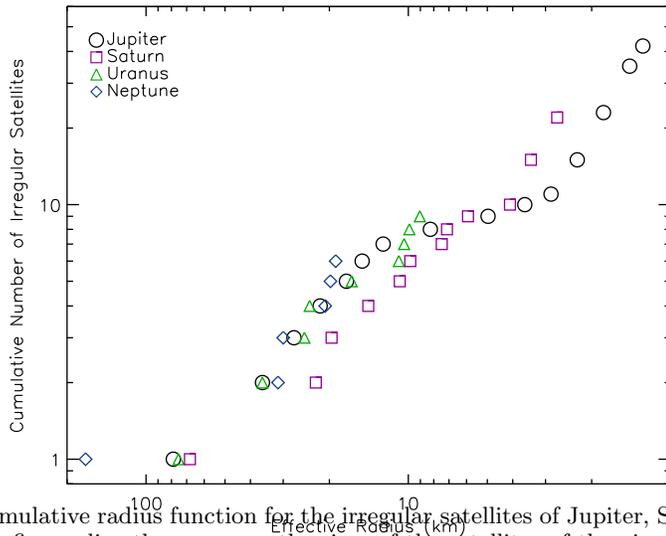}}
\caption{Cumulative radius function for the irregular satellites of
Jupiter, Saturn, Uranus and Neptune.  This figure directly compares
the sizes of the satellites of the giant planets assuming all
satellite populations have similar low albedos.  The planets have
statistically similar shallow size distributions of irregular
satellites.  Neptune's irregular satellite size distribution is
plotted without including Triton.  [Modified from Sheppard et
al. 2005]}
\label{fig:cum}
\end{figure}

When measured to a given size the population and size distributions of
the irregular satellites of each of the giant planets appears to be
very similar (Figure \ref{fig:cum}) (Sheppard and Jewitt 2005; Jewitt
and Sheppard 2005).  In order to model the irregular satellite size
distribution we use a differential power-law radius distribution of
the form $n(r)dr=\Gamma r^{-q}dr$, where $\Gamma$ and $q$ are
constants, $r$ is the radius of the satellite, and $n(r)dr$ is the
number of satellites with radii in the range $r$ to $r+dr$.  All giant
planet irregular satellite systems appear to have shallow power law
distribution of $q\sim 2$.  If we don't include Triton the largest
irregular satellite of each planet is of the 150 km scale with about
one hundred irregular satellites expected around each planet with
radii larger than about 1 km.  This is unexpected considering the
different formation scenarios envisioned for the gas giants versus the
ice giants.

\section{Discussion and Conclusions}

The irregular satellites of each planet are a distinct group of bodies
not necessarily linked to the two prominent reservoirs of the main
asteroid belt or the Kuiper Belt.  These satellites may have formed
relatively near their current locations and were subsequently captured
by their respective planet near the end of the planet formation epoch.
With the development of large, sensitive, digital detectors on large
class telescopes in the late 1990's the discovery and characterization
of the irregular satellites improved dramatically.  We find that the
gas giants Jupiter and Saturn and the ice giants Uranus and Neptune
all have a system of irregular satellites which have similar sizes,
populations and dynamics.

Current observations favor the capture mechanism of collisional or
collisionless interactions within the Hill spheres of the planets.
This capture mechanism is fairly independent of the planets formation
scenario and mass unlike gas drag or pull-down capture (Jewitt \&
Sheppard 2005).  Because the less massive ice giants are more distant
from the Sun their Hill spheres are actually larger than the gas
giants.  These increased Hill spheres may compensate for the lower
density of small bodies in the outer solar nebula and thus allow all
the giant planets to capture similar irregular satellite systems.
Recent discoveries of binaries in the Kuiper Belt show that such
objects may be quite common in the outer solar system.  These binary
pairs would be ideal for creating irregular satellites of the giant
planets through three body interactions as has been shown for the
capture of Triton (Agnor \& Hamilton 2004).  In fact, the equally
sized binary pairs in the Kuiper Belt may have formed in a similar
manner (Funato et al. 2004).

Three body interactions would have been much more probable in the
early solar system just after planet formation when leftover debris
was still abundant.  This capture process would allow for the possible
scattering predicted for Uranus and Neptune unlike gas drag and
pull-down capture since capture by three body interactions would still
operate after any scattering.  Three body capture also agrees with the
results of Beauge et al. (2002) in which they find the irregular
satellites would have to have formed after any significant planetary
migration as well as Brunini et al. (2002) who find that Uranus'
irregular satellites would have to be captured after any impact which
would have tilted the planet's rotation axis.  Also, Triton may have
disrupted the outer satellites of Neptune and capture of these
irregulars may have occurred after Triton was captured (Cuk \& Gladman
2005).  These scenarios all point to satellite capture happening just
after the planet formation process.

If three body interactions were the main capture mechanism then one
may expect the terrestrial planets to have irregular satellites.  The
terrestrial planets had very small Hill spheres compared to the giant
planets because of their low mass and proximity to the Sun.  In
addition, the terrestrial planets had no population of regular
satellites for passing objects to possibly interact with.  This may
explain why Mars and the other terrestrial planets have no outer
satellites, though Mars' two inner satellites may have been capture
through three body interactions. Perhaps Phobos and/or Deimos were
once binary asteroids.

The observed irregular satellite dynamical families were probably
created after capture.  In order to have a high probability of impact
for the creation of families either the captured had to occur very
early on when collisions were much more probable than now or there must
have been a much larger population of now defunct satellites around
each planet.

The non-detection of volatiles on Jupiter's irregular satellites
whereas volatiles are seen on Saturn's and Neptune's bodes well for
the objects to have formed near their current location.  The currently
limited data on the albedos, colors and densities of the irregular
satellites appear to show that each planet's irregular satellites are
physically distinct.  Jupiter's irregulars are remarkably similar to
the Jovian Trojans and dead comets.  Saturn's are significantly redder
but neither Jupiter's or Saturn's show the very red material observed
in the Kuiper Belt.  Uranus' irregulars have a wide range of colors
with some being the bluest and others being the reddest.

\begin{acknowledgments}
Support for this work was provided by NASA through Hubble Fellowship
grant \# awarded by the Space Telescope Science Institute, which is
operated by the Association of Universities for Research in Astronomy,
Inc., for NASA, under contract NAS 5-26555.
\end{acknowledgments}

\end{document}